\RequirePackage{ifpdf}
\documentclass[12pt,letterpaper]{JHEP3}         
\usepackage{amssymb,amsfonts}

\usepackage{epsfig}

\def\be{\begin{eqnarray}}
\def\ee{\end{eqnarray}}
\newcommand{\nn}{\nonumber}
\newcommand\para{\paragraph{}}

\newcommand{\eqn}[1]{(\ref{#1})}

\newcommand\zf{{\cal N}=(0,4)}
\newcommand\zt{{\cal N}=(0,2)}
\newcommand{\ff}{{\cal N}=(4,4)}
\newcommand{\adss}{AdS_3\times {\bf S}_+^3\times {\bf S}_-^3\times {\bf S}^1}
\newcommand{\pprime}{${}^\prime$}
\newcommand{\qfp}{Q_5^+}
\newcommand{\qfm}{Q_5^-}
\newcommand{\qo}{Q_1}

\def\Dslash{\,\,{\raise.15ex\hbox{/}\mkern-12mu D}}
\def\Dbarslash{\,\,{\raise.15ex\hbox{/}\mkern-12mu {\bar D}}}
\def\delslash{\,\,{\raise.15ex\hbox{/}\mkern-9mu \partial}}
\def\delbarslash{\,\,{\raise.15ex\hbox{/}\mkern-9mu {\bar\partial}}}
\def\pslash{\,\,{\raise.15ex\hbox{/}\mkern-9mu p}}
\def\calDslash{\,\,{\raise.15ex\hbox{/}\mkern-12mu {\cal D}}}

\newcommand{\Tr}{{\rm Tr}}

\def\lae{\mathrel{\mathop{\smash{\lower .5 ex \hbox{$\stackrel<\sim$}}}}}
\def\lae{\mathrel{\mathop{\smash{\lower .5 ex \hbox{$\stackrel>\sim$}}}}}

\title{The Holographic Dual of $AdS_3\times {\bf S}^3\times {\bf S}^3\times {\bf S}^1$}

\author{David Tong\\
Department of Applied Mathematics and Theoretical Physics, \\
University of Cambridge, UK\\
{\tt d.tong@damtp.cam.ac.uk}}

\abstract{We construct the two dimensional ${\cal N}=(0,4)$ gauge theory that lives on the world volume of D1-branes and intersecting D5-branes. We conjecture that this theory flows in the infra-red to a fixed point with large ${\cal N}=(4,4)$ 
superconformal symmetry. The central charge of the conformal field theory 
is shown to coincide with 
the holographic dual of string theory compactified on $AdS_3\times {\bf S}^3\times {\bf S}^3\times {\bf S}^1$.}

\begin{document}
\pagestyle{plain} \setcounter{page}{1}
\newcounter{bean}
\baselineskip16pt \setcounter{section}{0}

\section{Introduction and Summary}

Among the many  $AdS$ string compactifications, there is one which remains enigmatic. It is the solution of Type II string theory with geometry
\be AdS_3\times {\bf S}_+^3\times {\bf S}_-^3\times {\bf S}^1\label{geom}\ee
The geometry is supported by  NS fivebrane and string flux. In the Type IIB theory, there is also an S-dual configuration supported by RR fivebrane and string flux. The string charge of the configuration is denoted as $Q_1$, while each three sphere ${\bf S}^3_\pm$ is threaded with fivebrane flux $Q_5^\pm$. These fluxes, together with the string coupling constant, determine the radii of the ${\bf S}^3$ and ${\bf S}^1$.

\para
Strings propagating in the background \eqn{geom} were first studied in \cite{elitzur} and detailed discussions of the properties expected from the dual boundary field theory were given in \cite{jan,sergei}. 
The unusual feature of the geometry is the appearance of two ${\bf S}^3$ factors. This implies the existence of  two $SU(2)$ 
R-symmetries in the dual conformal field theory, each of which gives rise to a current algebra at level $k^\pm = Q_1Q_5^\pm$. This is the hallmark of a {\it large} ${\cal N}=(4,4)$ superconformal algebra. 

\begin{figure}[!h]
\begin{center}
\includegraphics[ width=3.2in,height=2.6in]{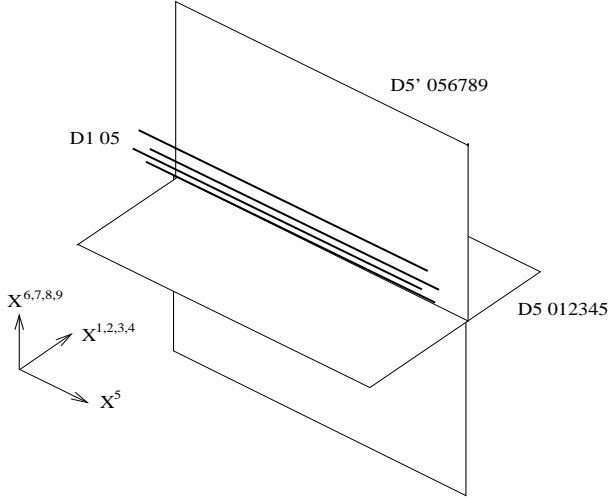}
\end{center}
\caption{The configuration of intersecting D5-branes and D1-branes}\label{ifig}
\end{figure}

\para
The existence of extended supersymmetry is usually a good thing. However, in the present case it appears to have been more of  a hinderance. The background \eqn{geom} is much less well understood that the $AdS_3\times {\bf S}^3\times {\cal M}$ geometries, with ${\cal M}=K3$ or $T^4$.  These correspond to small ${\cal N}=(4,4)$ superconformal theories, with just a single $SU(2)$ R-symmetry current, and can be studied through the familiar D1-D5 system.  In contrast,  no boundary field theory dual to the geometry \eqn{geom} has been identified.

\para
Of course, many features of the dual field theory can be extracted from a study of the supergravity solution. For our purposes, the most important is the central charge which can be computed by reducing on the spheres and circle and using the formula of \cite{bh}. The result  is given by \cite{sergei}\footnote{The central charge of a general large ${\cal N}=(4,4)$ SCFT with affine $\widehat{SU(2)}{}^\pm$ Lie algebras at levels $k^\pm$ takes the form $\hat{c} = 2k^+k^-/(k^++k^-)$ \cite{sevrin}.}
\be \hat{c} = 2Q_1 \frac{Q_5^+Q_5^-}{Q_5^++Q_5^-}\label{c}\ee
The purpose of this paper is to construct a boundary field theory whose infra-red fixed point reproduces this central charge.

\para
The simplest way to determine a dual boundary field theory is to build a D-brane configuration whose near-horizon limit gives the geometry of interest. The worldvolume dynamics of this D-brane configuration can often be identified with the dual field theory. For the geometry \eqn{geom}, there is a D-brane set-up that gets tantalisingly close. It consists of two stacks of D5-branes, with a number of D-strings lying at their intersection. The worldvolume direction of all branes is shown in the Figure \ref{ifig} and given by
\be \mbox{$Q_5^+$\ D5-branes}: && 012345\nn\\
\mbox{$Q_5^-$\ D5${}^\prime$-branes}: && 056789 \nn\\
 \mbox{$Q_1$\ D1-branes}: && 05\nn\ee
This configuration preserves ${\cal N}=(0,4)$ supersymmetry along the common $d=1+1$ worldvolume directions. 
A  corresponding supergravity solution exists, but only if  the D1-branes are smeared over the $x^{1,2,3,4}$ and $x^{6,7,8,9}$ directions. Taking the near-horizon limit, one finds the geometry \cite{cowdall,boonstra,jerome}
\be AdS_3\times {\bf S}_+^3\times {\bf S}_-^3\times {\bf R}\label{badgeom}\ee
The emergence of two ${\bf S}^3$ factors suggests that the low-energy dynamics of this configuration is governed by a theory with large ${\cal N}=(4,4)$ supersymmetry. However, the presence of the real line ${\bf R}$, rather than a circle ${\bf S}^1$ seen in \eqn{geom}, makes it difficult to interpret this geometry in terms of a boundary conformal field theory \cite{jan,sergei}.

\para
Nonetheless,  the main idea in this  paper is to take the D-brane configuration shown in Figure \ref{ifig} seriously as a candidate for the dual of the geometry \eqn{geom}. We start by constructing the ${\cal N}=(0,4)$ gauge dynamics that lives on the common worldvolume. The  D1-D1 strings, D1-D5 strings and D1-D5${}^\prime$ strings all give rise to vector multiplets and hypermultiplets that are familiar from the usual D1-D5 story, albeit with interactions that are twisted in an interesting manner. The  puzzle is how to deal with the D5-D5${}^\prime$ strings that lie at the intersection. These have 8 Dirichlet-Neumann directions and give rise to  left-moving chiral fermions. The obstacle to constructing this gauge theory in the past has been understanding how these chiral fermions couple to the other degrees of freedom.

\para
The answer to this puzzle is mildly surprising: the theory in the absence of the D5-D5${}^\prime$ strings does not preserve any supersymmetry.  In order to restore supersymmetry, we are obliged to include the D5-D5${}^\prime$ strings that are localised at the intersection, together with a very specific interaction term which couples them to  the D1-D5 and D1-D5\pprime \ strings. In this way, we can determine the ${\cal N}=(0,4)$ gauge theory that lives on the D-branes.

\para
Gauge theories in two dimensions are not conformal, but they can flow to interesting fixed points in the infra-red. In fact, with enough supersymmetry, gauge theories often flow to different, decoupled conformal field theories in the infra-red, associated to the  Higgs and Coulomb branches of the theory \cite{wittencomments,wittenhiggs}. For the ${\cal N}=(0,4)$ gauge theory on the D-branes, we will find that there is a conformal field theory that is not associated to the vacuum moduli spaces of the theory, but instead has  degrees of freedom localised at the intersection. The central charge can be determined by identifying an appropriate R-symmetry current in the infra-red. We show that this central charge is given by \eqn{c}.

\para
The agreement of the central charge provides evidence that the ${\cal N}=(0,4)$ gauge theory enhances its supersymmetry on the way down, flowing to the large ${\cal N}=(4,4)$ superconformal fixed point dual to $\adss$. This still leaves us with the question of how to interpret the factor of ${\bf R}$ in \eqn{badgeom} that arises in the near horizon limit. We suspect that this is simply a red herring, an artefact of supergravity most likely arising due to the smearing of D1-branes prior to taking the limit. If our proposal is correct, a supergravity solution describing localised branes would presumably give the required ${\bf S}^1$ factor instead. (See, however, \cite{plot1,plot2} for obstacles to constructing such localised solutions in various situations). 

\para
Throughout the paper, we will use the notation of ${\cal N}=(0,2)$ superfields. We start in Section \ref{sec1} by introducing these superfields and explaining how they fit together to form ${\cal N}=(0,4)$ multiplets. In Section \ref{dbranesec}, we describe the specific ${\cal N}=(0,4)$ gauge theory that lives on the intersection of D1, D5 and D5${}^\prime$-branes. In Section \ref{csec} we describe the relationship between the central charge and R-currents and use this to compute the central charge. 
 We close in Section \ref{endsec} with a number of open questions.

\section{${\cal N}=(0,4)$ Gauge Theories}\label{sec1}

In this section, we review a number of aspects of ${\cal N}=(0,4)$ multiplets and Lagrangians. We will construct these from ${\cal N}=(0,2)$ multiplets  and, for this reason, we begin with a review of ${\cal N}=(0,2)$ superfields. The presentation
follows \cite{phases} (similar presentations can also be found in \cite{abs,hetvort}). However, to keep formulae as simple as possible we will omit most numerical factors (usually $\sqrt{2}$ and $i$) on the grounds that they will not be important in the sequel.

\subsection{${\cal N}=(0,2)$}

Both ${\cal N}=(0,2)$ and ${\cal N}=(0,4)$ supersymmetry algebras have the characteristic feature that they contain only right-moving supercharges. This means that right-handed fermions are paired with bosonic fields in the familiar supersymmetric manner. In contrast, left-moving fermions can be loners; they need not have  bosonic companions, in which case supersymmetry acts only to restrict their interactions with the  right-handed fermions.

\para

${\cal N}=(0,2)$ superspace in $d=1+1$ dimensions is parameterised by the spacetime coordinates, $x_\pm = x^0\pm x^1$, and a single, complex right-moving Grassmann parameter $\theta^+$. The chiral supercharges $Q_+$ and $\bar{Q}_+$ are defined by
\be
Q_+=\frac{\partial}{\partial\theta^+}+i\bar{\theta}^+(\partial_0+\partial_1) \ \ \ {\rm and}\ \ \ 
\bar{Q}_+=-\frac{\partial}{\partial\bar{\theta}^+}-i\theta^+(\partial_0+\partial_1) \nn
\ee
Theories with ${\cal N}=(0,2)$ supersymmetry have a  $U(1)_R$ R-symmetry under which $\theta^+$ has charge $+1$. This $U(1)_R$ symmetry will play a key role in Section \ref{csec} where we use it to compute the central charge of our theory.

\para
We will be interested in three types of superfields: gauge, chiral and Fermi multiplets. We deal with each in turn.

\subsubsection*{Gauge Multiplets}

The gauge multiplet lives in a real superfield, $U$. It comprises of a gauge field $u$, an adjoint-valued left-handed fermion $\zeta_-$ and a real auxiliary field $D$. The  component expansion is
\be
U=(u_0-u_1)-i\theta^+\bar{\zeta}_- - i\bar{\theta}^+\zeta_-+\theta^+\bar{\theta}^+D \nn\ee 
As is common in supersymmetric theories, in order to write down the action for the gauge field we first define a new multiplet which contains the field strength. To do this, we need to introduce some covariant super-derivatives. The right-handed derivatives act on superspace and are defined by
\be {\cal D}_+=\frac{\partial}{\partial
\theta^+}-i\bar{\theta}^+({\cal D}_0+{\cal D}_1)
\ \ \ \ {\rm and}\
\ \ \ \ \bar{{\cal D}}_+=-\frac{\partial}{\partial
\bar{\theta}^+}+i{\theta}^+({\cal D}_0+{\cal D}_1)\nn\ee
where $ {\cal D}_0+{\cal D}_1=\partial_0+\partial_1-i(u_0+u_1)$
includes the $u_+$ component of the gauge field, but no fermions. Meanwhile, the left-handed derivative does not include a derivative on superspace, but does include fermions. It   is given by 
\be {\cal D}_-= {\cal D}_0-{\cal D}_1=\partial_0-\partial_1-iU \nn\ee
The field strength, $u_{01}=\partial_0u_1-\partial_1u_0-i[u_0,u_1]$, then lives in a multiplet constructed from the commutator of derivatives, 
\be \Upsilon=[\bar{\cal D}_+,{\cal D}_-]=- \zeta_--i\theta^+(D-iu_{01})-i\theta^+\bar{\theta}^+({\cal
D}_0+{\cal D}_1)\zeta_-\nn\ee
It turns out that $\Upsilon$ is an example of a Fermi multiplet; we shall give the more general definition below.

\para
The standard action for the gauge multiplet is written as
\be S_{\rm gauge} &=& \frac{1}{8g^2}\,\Tr\,\int d^2x \,d^2\theta\ \Upsilon^\dagger \Upsilon 
 \nn \\&=&\frac{1}{g^2}\,\Tr\,\int d^2x\ \left(\frac{1}{2} u_{01}^2+i\bar{\zeta}_-({\cal D}_0+{\cal D}_1)\zeta_-+D^2\right)
\label{gaugelag}\ee

\subsubsection*{Chiral Multiplets}

${\cal N}=(0,2)$ chiral multiplets contain a right-moving fermion $\psi_+$ and a single complex scalar $\phi$, each transforming in the same representation of the gauge group. (For our purposes, we will need only fundamental and adjoint representations).  These sit together in a complex-valued bosonic superfield $\Phi$ obeying 
\be \bar{\cal D}_+\Phi=0 .\ee
The component expansion is
\be
\Phi=\phi+ \theta^+\psi_+-i\theta^+\bar{\theta}^+(D_0+D_1)\phi
\label{02chi}\ee
where $(D_0+D_1)$ is the usual bosonic covariant derivative. A simple, but important, observation: if $\phi$ has R-charge $R[\phi]$, then $\psi_+$ has R-charge $R[\psi_+]=R[\phi]-1$.

\para
The kinetic terms for the chiral multiplet are given by the
action,
\be S_{\rm chiral}&=& \int d^2x\,d^2\theta\
\bar{\Phi}({\cal D}_0-{\cal D}_1)\Phi \nn \\ &=& \int
d^2x\ \left(-|D_\mu\phi|^2+i\bar{\psi}_+(D_0-D_1)\psi_+ -
i\bar{\phi}\zeta_-\psi_+ +
i\bar{\psi}_+\bar{\zeta}_-\phi + \bar{\phi}D\phi\right) \ \ \ \ \ \label{chirallag}\ee

\subsubsection*{Fermi Multiplets}

As we mentioned above, ${\cal N}=(0,2)$ theories have the property that left-moving fermions are not necessarily accompanied by propagating, bosonic superpartners. A fermion of this kind, $\psi_-$, sits in a Fermi multiplet $\Psi$.  They obey the condition 
\be \bar{\cal D}_+\Psi= E(\Phi_i)\label{e}\ee
where $E(\Phi_i)$ is some holomorphic function of the chiral superfields $\Phi_i$. This function must be chosen so that $E(\Phi_i)$ transforms in the same manner as $\Psi$ under any symmetries; examples will be given below.  
The choice of $E$ determines the interaction of the theory and also appears in the component expansion of the superfield,
\be \Psi = \psi_-- \theta^+ G
-i\theta^+\bar{\theta}^+(D_0+D_1)\psi_--\bar{\theta}^+E(\phi_i) +\theta^+\bar{\theta}^+\frac{\partial E}{\partial \phi^i}\psi_{+i}
\label{02fer}\ee
Here $G$ is a complex auxiliary field. The superfield $\Upsilon$ which contains the field strength is of this type, with 
 $\bar{\cal D}_+\Upsilon = 0$.
 
 \para 
  The kinetic terms for the fermi multiplet are  given by
\be S_{\rm fermi}&=& \int d^2x\,d^2\theta\
\bar{\Psi}\Psi \nn \\&=&
\int d^2x\ i\bar{\psi}_-(D_0+D_1)\psi_-+|G|^2-|E(\phi_i)|^2
-\bar{\psi}_-\frac{\partial E}{\partial\phi^i}\psi_{+i}
+\bar{\psi}_{+i}\frac{\partial\bar{E}}{\partial\bar{\phi}_i}\psi_-\ \ \ \ \ \ \ \ 
\label{fermilag}\ee
Note that the holomorphic function $E(\phi_i)$ appears as a potential term in the Lagrangian. However, as we now explain, 
this is not the only way to introduce potential terms for chiral multiplet scalars.

\subsection*{Superpotentials}

Each Fermi multiplet contains an auxiliary  complex scalar $G$. We can use this to construct further potential terms for chiral multiplets. To do this,  we introduce a holomorphic function $J^a(\Phi_i)$ for each Fermi multiplet $\Psi_a$. We  can then build a supersymmetric action by integrating over half of superspace, 
\be S_{J} =  \int d^2x\,d\theta^+ \ \
\sum_a\Psi_a 
J^a(\Phi_i)\ +\ {\rm h.c.}\label{dohalf}\ee
where the integrand is implicitly evaluated at $\bar{\theta}^+=0$. This yields the following action
\be
S_J = \sum_{a}\int d^2x\ \
G_aJ^a(\phi_i)+\sum_i\psi_{-a}\frac{\partial
J^a}{\partial\phi_i}\psi_{+i}+\ {\rm h.c.}\label{sj}\ee
After integrating out the auxiliary fields $G_a$, this results in a potential term $\sim |J^a(\phi_i)|^2$. This is usually referred to as the  {\it superpotential} in ${\cal N}=(0,2)$ theories. In the following, we  will also use the notation
\be {\cal W} = \Psi_aJ^a(\Phi)\nn\ee
We see that there are two ways to construct potential terms in theories with ${\cal N}=(0,2)$ supersymmety. Both are associated to Fermi multiplets and both involve holomorphic functions, $E(\phi_i)$ and $J(\phi_i)$. The difference between them is not visible in the bosonic Lagrangian alone; it only shows through the subtle difference in the Yukawa terms in \eqn{fermilag} and \eqn{sj}. This difference will be important when we come to discuss ${\cal N}=(0,4)$ theories next. 

\para
There is one final condition that is necessary  to ensure that our actions are indeed invariant under ${\cal N}=(0,2)$ supersymmetry. The integration over half of superspace in \eqn{dohalf} is supersymmetric if and only if 
$\bar{D}_+(\Psi_aJ^a)=0$. This requires 
\be E\cdot J \equiv \sum_a E_aJ^a=0 \label{ej} .\ee
This shows that there is some tension when introducing both $E$-type potentials and $J$-type potentials associated to the same Fermi multiplet. In general, this is possible only if we can arrange some cancellation between other Fermi multiplets. 
This condition will prove crucial in our construction of $\zf$ theories below. 

\subsection{${\cal N}=(0,4)$}\label{04sec}

Theories with ${\cal N}=(0,4)$ supersymmetry have four, real, right-moving supercharges. Correspondingly, they have an R-symmetry
\be SO(4)_R \cong SU(2)^-_R\times SU(2)^+_R\nn\ee
The supercharges transform in the $({\bf 2},{\bf 2})_+$ representation, where the subscript denotes their chirality under the $SO(1,1)$ Lorentz group (and is not to be confused with the $\pm$ superscript on the R-symmetry groups).

\para
Our goal in this section is to describe the multiplet structure and Lagrangians for gauge theories with ${\cal N}=(0,4)$ supersymmetry.  Our strategy is to build theories using ${\cal N}=(0,2)$ supermultiplets which enjoy an enhanced $SO(4)_R$ R-symmetry, ensuring that there is an extended supersymmetry. As we now describe, there are four basic multiplets with ${\cal N}=(0,4)$ supersymmetry: vector multiplets, hypermultiplets, twisted hypermultiplets and Fermi multiplets.

\subsubsection*{Vector Multiplets}

The $\zf$ vector multiplet is comprised of an $\zt$ vector multiplet $U$ together with an adjoint-valued $\zt$ Fermi multiplet $\Theta$. As well as the gauge field, there are a pair of left moving complex fermions, $\zeta^a_-$,  $a=1,2$, transforming as $({\bf 2},{\bf 2})_-$ under the R-symmetry and a triplet of auxiliary fields transforming as $({\bf 3},{\bf 1})$.

\para
The Fermi multiplet obeys
\be \bar{\cal D}_+\Theta= E_\Theta\label{etheta}\ee
where the function $E_\Theta$ depends on the matter content and will be discussed further below. The $\zf$ vector multiplet Lagrangian is given by the sum of the \eqn{gaugelag} and \eqn{fermilag}. 

\subsubsection*{Hypermultiplets}

There are two distinct ways to couple matter fields to a gauge multiplet. For this reason, we distinguish between hypermultiplets and twisted hypermultiplets. 

\para
An $\zf$ hypermultiplet consists of a pair of $\zt$ chiral multiplets, $\Phi$ and $\tilde{\Phi}$, transforming in conjugate representations of the gauge group. The pair of complex scalars transforms as $({\bf 2},{\bf 1})$ under the R-symmetry, while the pair of right-moving fermions transforms as $({\bf 1},{\bf 2})_+$. 

\para
The kinetic terms for both chiral multiplets are given by \eqn{chirallag}. In addition, there is a coupling to Fermi field $\Theta$ in the ${\cal N}=(0,4)$ vector multiplet. This takes the form of a superpotential \eqn{sj} with
\be J^\Theta = \Phi\tilde{\Phi}\ \ \ \ \Rightarrow \ \ \ \ \ {\cal W}_\Theta = \tilde{\Phi}\Theta\Phi \label{zfsup}\ee
This is entirely analogous to the superpotential needed to construct ${\cal N}=(4,4)$ theories from ${\cal N}=(2,2)$ superfields. Indeed, if one decomposes an ${\cal N}=(4,4)$ hypermultiplet, the bosonic fields and right-moving fermions sit in an  $\zf$ hypermultiplet, coupled via the  superpotential \eqn{zfsup}.

\subsubsection*{Twisted Hypermultiplets}

The $\zf$ twisted hypermultiplet also consists of a pair of $\zt$ chiral multiplets, $\Phi'$ and $\tilde{\Phi}'$, transforming in conjugate representations of the gauge group. The difference from the hypermultiplet lies in the R-symmetry transformation of the fields. The pair of scalars transform as $({\bf 1},{\bf 2})$ while the pair of right-moving fermions transforms as $({\bf 2},{\bf 1})_+$.

\para
The kinetic terms for the two chiral multiplets are again given by \eqn{chirallag}. The different R-symmetry transformations are enforced by the coupling to the Fermi field $\Theta$.  In contrast to the hypermultiplet, the coupling is no longer through the superpotential but instead via the relation \eqn{etheta}, with 
\be E_\Theta =  \Phi'\tilde{\Phi}'\label{esaregood}\ee
with the combination $\Phi'\tilde{\Phi}'$ arranged so that it transforms in the adjoint of the gauge group.

\para
Perhaps the most familiar example of a twisted hypermultiplet arises in the  construction of the ${\cal N}=(4,4)$ theories. The ${\cal N}=(4,4)$ vector multiplet consists of an ${\cal N}=(0,4)$ vector multiplet, together with an adjoint-valued ${\cal N}=(0,4)$ twisted hypermultiplet.  

\para
To preserve $\zt$ supersymmetry, we must have $E\cdot J=0$. If we have only a single vector multiplet, this condition reads
\be E_\Theta J^\Theta = 0\nn\ee
Obviously this is not satisfied if we naively try to couple both hypermultiplets and twisted hypermultiplets to the same gauge group. To do this in  a manner consistent with supersymmetry, we need to introduce further multiplets. These are:

\subsubsection*{Fermi Multiplets}

\para

We define the $\zf$ Fermi multiplets to consist of a pair of $\zt$ Fermi multiplets, $\Gamma$ and $\tilde{\Gamma}$, transforming in conjugate representations of the gauge group. The left-handed fermions transform as $({\bf 1},{\bf 1})_-$ under the R-symmetry.

\para
The kinetic terms for each fermion are given by \eqn{fermilag}; no further coupling between $\Gamma$, $\tilde{\Gamma}$ and $\Theta$ is needed (nor, indeed, possible, since both contain left-moving fermions only). It is, however, possible to introduce other couplings for the Fermi multiplet through the potentials $E$ and $J$ in a manner that preserves $SO(4)_R$ R-symmetry. While we have not determined the most general such interaction, the one that will be relevant for our purposes couples a Fermi multiplet to a hypermultiplet and a twisted hypermultiplet. It arises, as we  explain shortly, when we look at the better studied $\ff$ gauge theories through $\zf$ eyes.

\para
Finally, we note that it is possible to have a single ${\cal N}=(0,2)$ Fermi multiplet which is consistent with ${\cal N}=(0,4)$ supersymmetry. This possibility was pointed out in \cite{wadhm,putrov}. For this to happen, the chiral fermion should be a singlet under the $SO(4)_R$ symmetry. Of course, the coupling to other matter multiplets must also respect this. As we will see below, the D5-D5' string provide an example of this.

\subsubsection*{${\cal N}=(4,4)$ Decomposition}

To gain some familiarity with ${\cal N}=(0,4)$ theories, it will prove useful to see how ${\cal N}=(4,4)$ multiplets decompose into their $\zf$ counterparts.

\para
The $\ff$ vector multiplet splits into an $\zf$ vector multiplet and an adjoint-valued $\zf$ twisted hypermultiplet. We will denote the $\zt$ chiral multiplets in this twisted hypermultiplet as $\Sigma$ and $\tilde{\Sigma}$. They couple to the vector mutliplet $\Theta$ through,
\be E_\Theta =  [\Sigma,\tilde{\Sigma}]\nn\ee
An $\ff$ hypermultiplet decomposes into an $\zf$ hypermultiplet,  $\Phi$ and $\tilde{\Phi}$, and an $\zf$ Fermi multiplet, $\Gamma$ and $\tilde{\Gamma}$. As described above, there is a superpotential term 
\be {\cal W}_\Theta = \tilde{\Phi}\Theta\Phi\nn\ee
This can be traced to the more familiar superpotential term that arises when writing $\ff$ theories in terms of ${\cal N}=(2,2)$ superfields.

\para
The remaining couplings are associated to the $\zf$ Fermi multiplet and provide the example of an interaction between a Fermi multiplet $\Gamma,\tilde{\Gamma}$, a hypermultiplet $\Phi$,$\tilde{\Phi}$ and a twisted hypermultiplet $\Sigma$,$\tilde{\Sigma}$ that we promised above. The interaction makes use of both superpotentials and $E$ terms. The former are given by
\be {\cal W}_{\tilde{\Gamma}} +{\cal W}_{\Gamma} = \tilde{\Gamma}\tilde{\Sigma}\Phi + \tilde{\Phi}\tilde{\Sigma}\Gamma \label{44super}
\ee
These are combined with  the E-term couplings, 
\be E_\Gamma = \Sigma\Phi\ \ \ \ ,\ \ \ \ E_{\tilde{\Gamma}} = -\tilde{\Phi}\Sigma
\label{44e}\ee
Note that the constraint \eqn{ej} is satisfied, as required, because
\be E\cdot J  = \left(\tilde{\Phi}[\Sigma,\tilde{\Sigma}]\Phi + \tilde{\Phi}\tilde{\Sigma}\Sigma\Phi-\tilde{\Phi}\Sigma \tilde{\Sigma}\Phi\right)=0\nn\ee
In the next section, we will see how to construct  a gauge theory on D-branes which exhibits only $\zf$ supersymmetry. Nonetheless, it will be constructed from the $\ff$ ingredients that we described above.

\section{D-Brane Dynamics}\label{dbranesec}

Our goal in this section is to determine the dynamics of the following configuration of D-branes:
\be \mbox{$Q_5^+$\ D5-branes}: && 012345\nn\\
\mbox{$Q_5^-$\ D5${}^\prime$-branes}: && 056789 \nn\\
 \mbox{$Q_1$\ D1-branes}: && 05\nn\ee
The D5 and D5\pprime-branes alone preserve  ${\cal N}=(0,8)$ supersymmetry; the D1-branes break this to $\zf$. We are interested in the worldvolume theory on their shared $d=1+1$ intersection. This has  $U(Q_1)$ gauge group, with $SU(\qfm)\times SU(\qfp)$ both arising as global symmetries. In addition, there is a $G\cong SO(4)^-\times SO(4)^+$ symmetry arising from spatial rotations, which we write as 
\be  G \cong SU(2)^-_L\times SU(2)^-_R\times SU(2)^+_L\times SU(2)^+_R\nn\ee
Of these, $SU(2)^-_R\times SU(2)^+_R$ is the R-symmetry group shared by all $\zf$ theories. The $SU(2)^-_L\times SU(2)^+_L$ is a global symmetry.

\para
We will look at the multiplets we get  from quantising  fundamental strings attached to different kinds of branes.

\para
\underline{D1-D1 Strings}
\para
The low-energy dynamics of the D1-branes is the well-studied ${\cal N}=(8,8)$ $U(\qo)$ gauge theory. In $\zf$ language, the multiplets, their components and transformation under $G$ are
\be \mbox{Vector Multiplet}&:&\ \ \ \ (\zeta_-,\tilde{\zeta}_-)\  \ \ \ \ ({\bf 1},{\bf 2},{\bf 1},{\bf 2})\nn\\
   \mbox{Twisted Hypermultiplet}&:&\ \ \ \  (Y,\tilde{Y}^\dagger)\  \ \ \ \ \ ({\bf 1},{\bf 1},{\bf 2},{\bf 2})\nn\\
&:&\ \ \ \ (\zeta_+,\tilde{\zeta}^\dagger_+)\  \ \ \ \ ({\bf 1},{\bf 2},{\bf 2},{\bf 1}) \nn\\
    \mbox{Hypermultiplet}  &:&\ \ \ \ (Z,\tilde{Z}^\dagger)\  \ \ \ \ ({\bf 2},{\bf 2},{\bf 1},{\bf 1})\nn\\
&:&\ \ \ \ (\lambda_+,\tilde{\lambda}^\dagger_+)\  \ \ \ ({\bf 2},{\bf 1},{\bf 1},{\bf 2}) \nn\\
       \mbox{Fermi Multiplet}&:&\ \ \ \ (\lambda_-,\tilde{\lambda}_-)\  \ \ \ ({\bf 2},{\bf 1},{\bf 2},{\bf 1}) \nn\ee
Each of these multiplets transforms in the adjoint  of the $U(Q_1)$ gauge group. The twisted hypermultiplet scalars, $Y$ and $\tilde{Y}$, parameterise the directions $x^{6,7,8,9}$ parallel to the D5$'$-branes; the hypermultiplet scalars, $Z$ and $\tilde{Z}$, parameterise the directions $x^{1,2,3,4}$ parallel to the D5-branes.

\para
The $\zf$ hypermultiplet is charged under the vector multiplet, so necessarily couples through  the superpotential \eqn{zfsup}. Similarly, the $\zf$ twisted hypermultiplet is also charged under the gauge group so necessarily couples through the $E$-term \eqn{esaregood}. There are further couplings between these fields which are necessary to give rise to ${\cal N}=(8,8)$ supersymmetry. There is a superpotential term involving the Fermi multiplet which, in $\zt$ notation, we call $\Lambda$ and $\tilde{\Lambda}$,
\be {\cal W} =  {\rm Tr}\,\left(\Lambda\,[\tilde{Y},\tilde{Z}] + \tilde{\Lambda}\,[\tilde{Y},Z]\right) 
\nn\ee
In addition, there is the E-term coupling
\be E_\Lambda = [Y,Z] \ \ \ ,\ \ \ \ E_{\tilde{\Lambda}} =  [ Y,\tilde{Z}]\nn\ee

\para
\underline{D1-D5 Strings}
\para

The strings stretched between the D1-branes and D5-branes give rise to $\ff$ hypermultiplets. In $\zf$ notation, this becomes
\be   \mbox{Hypermultiplet}  &:&\ \ \ \ (\phi,\tilde{\phi}^\dagger)\  \ \ \ \ \ \ ({\bf 1},{\bf 2},{\bf 1},{\bf 1})\nn\\
&:&\ \ \ \ (\psi_+,\tilde{\psi}^\dagger_+)\  \ \ \ ({\bf 1},{\bf 1},{\bf 1},{\bf 2}) \nn\\
       \mbox{Fermi Multiplet}&:&\ \ \ \ (\psi_-,\tilde{\psi}_-)\  \ \ \ ({\bf 1},{\bf 1},{\bf 2},{\bf 1}) \nn\ee
Each of these multiplets transforms as ${\bf Q}_1$ under the $U(Q_1)$ gauge group and $\bar{\bf Q}_5^+$ under the global $SU(\qfp)$.  This means, in particular, that the hypermultiplet couples to the $U(Q_1)$ vector multiplet fields through the superpotential \eqn{zfsup}. 

\para
There is also a coupling between these strings and the adjoint twisted hypermultiplet, consisiting of $Y$ and $\tilde{Y}$. This reflects the fact that these strings become massive if the D1-branes move in the $x^{6,7,8,9}$ directions. The coupling takes the form of the superpotential \eqn{44super}
\be {\cal W} = \tilde{\Psi}\tilde{Y}\Phi + \tilde{\Phi}\tilde{Y}\Psi \nn
\nn\ee
together with the E-terms \eqn{44e} for the D1-D5 Fermi multiplet, 
\be E_\Psi =   Y\Phi\ \ \ \ ,\ \ \ \ E_{\tilde{\Psi}} = -\tilde{\Phi} Y
\nn\ee
The theory we have described above is  simply the well-studied D1-D5 system  in $\zf$ notation.  We now add the D5$'$-branes into the mix. 

\para
\underline{D1-D5$'$ Strings}
\para

Clearly, the strings  stretched between the D1-branes and D5$'$-branes must give rise to the same matter content as those stretched between the D1-branes and D5-branes. The only difference is that the $SO(4)^-$ and $SO(4)^+$ symmetries are exchanged. From our discussion of $\zf$ multiplets, this means that we now have a twisted hypermultiplet. The component fields are
\be   \mbox{Twisted Hypermultiplet}  &:&\ \ \ \ (\phi',\tilde{\phi}^{\prime\,\dagger})\  \ \ \ \ \ ({\bf 1},{\bf 1},{\bf 1},{\bf 2})\nn\\
&:&\ \ \ \ (\psi^\prime_+,\tilde{\psi}^{\prime\,\dagger}_+)\  \ \ \ \,({\bf 1},{\bf 2},{\bf 1},{\bf 1}) \nn\\
       \mbox{Fermi Multiplet}&:&\ \ \ \ (\psi^\prime_-,\tilde{\psi}^\prime_-)\  \ \ \ \ ({\bf 2},{\bf 1},{\bf 1},{\bf 1}) \nn\ee
Each of these multiplets transforms in the ${\bf Q}_1$ of the gauge group and $\bar{\bf Q}_5^-$ of the global $SU(\qfm)$. This means, in particular, that the twisted hypermultiplet couples to the $U(Q_1)$ vector multiplet through the E-term \eqn{esaregood}.  There must also be a further coupling between these fields and the adjoint hypermultiplet consisting of $Z$ and $\tilde{Z}$, reflecting the fact that these strings become massive when the D1-branes move in the $x^{1,2,3,4}$ directions.  These take the form of a superpotential,  
\be {\cal W} = \tilde{\Psi}'\tilde{Z}\Phi' + \tilde{\Phi}'\tilde{Z}\Psi' \nn
\nn\ee
together with the E-terms
\be E_{\Psi'} =   Z\Phi'\ \ \ \ ,\ \ \ \ E_{\tilde{\Psi}'} = -\tilde{\Phi}' Z
\nn\ee
One might naively think that this is the end of the story. In fact, it cannot be because the interactions that we have described so far are not supersymmetric! The trouble arises because of the constraint $E\cdot J=0$ given in \eqn{ej}. For the interactions described above, we have
\be E\cdot J &=& {\rm Tr}\,([Y,\tilde{Y}]+\Phi'\tilde{\Phi}')([Z,\tilde{Z}]+\Phi\tilde{\Phi}) + {\rm Tr}\,([Y,\tilde{Z}][\tilde{Y},Z]-[Y,Z][\tilde{Y},\tilde{Z}])\nn\\ && + (\tilde{\Phi}\tilde{Y}Y\Phi - \tilde{\Phi}Y\tilde{Y}\Phi)
+(\tilde{\Phi}'\tilde{Z}Z\Phi' - \tilde{\Phi}'Z\tilde{Z}\Phi') 
\nn\ee
where the first terms come from the vector multiplet contribution $E_\Theta J^\Theta$. The remaining terms come from $\Lambda$,$\tilde{\Lambda}$ and $\Psi$,$\tilde{\Psi}$ and  $\Psi'$,$\tilde{\Psi}'$ respectively. The sum of these does not vanish. Instead, we have
\be E\cdot J = {\rm Tr}\,(\Phi'\tilde{\Phi}'\Phi\tilde{\Phi})\label{remnant}\ee
In order to describe a supersymmetric theory, we must introduce further Fermi multiplets to cancel this contribution. Happily, such multiplets can be found, nestling in the heart of our D-brane configuration.

\para
\underline{D5-D5$'$ Strings}
\para

Quantising the strings that lie at the intersection of the two stacks of D5-branes gives to $Q_5^+Q_5^-$ chiral fermions which we collectively denote as $\chi_-$. (To see this, one can note, for example,  that the D5-branes can be T-dualised to the D0-D8 system described in \cite{evanati}). These sit in a single ${\cal N}=(0,2)$ Fermi multiplet\footnote{I'm grateful to Chi-Ming Chang and Shu-Heng Shao for pointing this out, correcting an error in an earlier version of this paper.},
\be  \mbox{${\cal N}=(0,2)$ Fermi Multiplet}  &:&\ \ \ \ \chi_-\  \ \ \ \ ({\bf 1},{\bf 1},{\bf 1},{\bf 1}) \nn\ee
%
%
Each of these fields transforms as a singlet under the $U(Q_1)$ gauge group, but transforms in the $({\bf Q}_5^-,\bar{\bf Q}_5^+)$ representation of the $SU(\qfm)\times SU(\qfp)$ global symmetry. To construct our $\zf$ D-brane theory, we need a to couple these Fermi mutliplets to the D1-D5 and D1-D5$'$ strings a manner compatible with these quantum numbers, preserving the spacetime symmetry $G$ and cancelling the $E\cdot J$ remnant \eqn{remnant}. This is achieved by
\be {\cal W} = \tilde{\Phi}\Phi'{\chi}\ \ \ \ {\rm and}\ \ \ \ E_\chi = -\tilde{\Phi}'\Phi\label{impey}\ee
(An alternative choice of ${\cal W} = \tilde{\Phi}'\Phi{\chi}$ and $E_\chi = -\tilde{\Phi}\Phi'$ differs from this by $\chi_-\rightarrow \bar{\chi}_-$.) The effect of these interactions is to ensure that
%
%
%
%
%
%
%
%
%
%
%
\be E\cdot J = 0\nn\ee
hence restoring supersymmetry.

\subsection{The Scalar Potential}\label{vacsec}

The scalar potential of the D-brane theory consists of the sum of $D$-terms, G-terms and $E$-terms in the interactions described above. It's instructive to gather these terms together in a manner which makes the full symmetry of the theory manifest. To this end, we introduce the notation
\be \omega = \left(\begin{array}{c} \phi \\ \tilde{\phi}^\dagger\end{array}\right)\ \ \ \ {\rm and}\ \ \ \ \ \omega' = \left(\begin{array}{c} \phi' \\ \tilde{\phi}'^\dagger\end{array}\right)\nn\ee
The doublet $\omega$ transforms in the ${\bf 2}$ of $SU(2)^-_R$ and $\omega'$ in the ${\bf 2}$ of $SU(2)^+_R$. Using this we construct  triplets of ``D-terms", each transforming in the adjoint of $U(Q_1)$,
\be \vec{D}_Z= \vec{\eta}_{ij}Z^iZ^j + \omega^\dagger \vec{\sigma}\omega\ \ \ \ {\rm and}\ \ \ \ \  \vec{D}_Y = \vec{\eta}_{ij}Y^iY^j + \omega^{\prime\,\dagger} \vec{\sigma}\omega'\nn\ee
Here $\vec{\sigma}$ are the Pauli matrices and $\vec{\eta}$ are the self-dual 't Hooft matrices, and we've indulged in a slight abuse of our previous notation because now $Z^i$, with $i=1,2,3,4$ denotes the four adjoint scalars $X^{1,2,3,4}$ while $Y^i$ denotes $X^{6,7,8,9}$. These are designed so that $\vec{D}_Z$ transforms in the ${\bf 3}$ of $SU(2)_R^-$ and $\vec{D}_Y$ transforms in the ${\bf 3}$ of $SU(2)_R^+$. Both are singlets under $SU(2)_L^-\times SU(2)_L^+$.

\para
These are very similar to the triplet of D-terms that arise  in the usual D1-D5 story with ${\cal N}=(4,4)$ supersymmetry. The only difference is that we now have a pair of them. The scalar potential is 
\be V = {\rm Tr}\,(\vec{D}_Z\cdot \vec{D}_Z + \vec{D}_Y\cdot \vec{D}_Y) + \omega^\dagger Y^iY^i \omega + \omega^{\prime\,\dagger}Z^iZ^i\omega' + {\rm Tr}[Y^i,Z^j]^2+ {\rm Tr}\,\left(\omega^\dagger\cdot \omega\,\omega^{\prime\,\dagger}\cdot\omega^{\prime}\right)\nn\ee
The last of these terms is perhaps the least familiar. It arises from the couplings  \eqn{impey}, together with $\omega\omega'$ cross-terms that arose from the original D-terms but are not included in $\vec{D}_Z$ and $\vec{D}_Y$ above. The indices in the bracket are constructed to be singlets under $SO(4)_R$ and the flavour groups, but adjoint under the $U(Q_1)$ gauge group. In more detailed notation, it is given by
\be
{\rm Tr}\,\left(\omega^\dagger\cdot \omega\,\omega^{\prime\,\dagger}\cdot\omega^{\prime}\right) = 
\sum_{a=1}^{Q_5^+}\sum_{b=1}^{Q_5^-}&&\left((\phi_a^\dagger\phi_b')(\phi_b^{\prime\,\dagger}\phi_a)
+(\tilde{\phi}_a\tilde{\phi}_b^{\prime\,\dagger})(\tilde{\phi}_b^{\prime} \tilde{\phi}_a^\dagger) 
\right. \nn\\ &&\ \ \ \ \ \ \ \ \ \ 
+ \left.(\phi_a^\dagger\tilde{\phi}_b^{\prime\,\dagger}) (\tilde{\phi}_b^\prime\phi_a)+(\tilde{\phi}_a\phi^\prime_b)(\phi_b^{\prime\,\dagger}\tilde{\phi}_a^\dagger)\right)
\nn\ee

\para
The potential $V$ has three distinct branches of vacua. 
\begin{itemize}
\item $\omega =\omega'=0$ with $Z^i$ and $Y^i$ mutually commuting.
\item $\vec{D}_Z=0$ and $Y^i=\omega'=0$
\item $\vec{D}_Y=0$ and $Z^i=\omega=0$
\end{itemize}
The first is the Coulomb branch and corresponds to the D-strings living away from the two stacks D5-branes. The second and third are both Higgs branches in which the D-strings are absorbed in {\it either} the D5-branes or the D5\pprime-branes. 
There are also mixed branches where different D-strings do their own thing.

\para
In $\ff$ theories, each branch of vacua is associated with a different conformal field theory in the infra-red \cite{wittenhiggs}.  In the present case, we claim that the CFT of interest is not associated to either of these branches, but instead is localised at the intersection $Z^i=Y^i=0$. 
We now provide  evidence for this claim.

\section{The Central Charge}\label{csec}

In this section we describe how to compute the central charge of our theory. We start by reviewing the computation of 
the central charge of $\zt$ theories in terms of the R-current anomaly. We then explain the application of this method 
to our $\zf$ theory. 

\para
In ${\cal N}=(0,2)$ superconformal theories the right-moving R-current $R$ sits in the same multiplet as the holomorphic stress tensor.  The operator product expansion of the currents includes a singular term that is proportional to the right-moving central charge, $\hat{c}_R={c}_R/3$. This means that the R-current anomaly is related to the central charge, 
\be \hat{c}_R= \Tr\,R^2\label{crawley}\ee
where $\Tr$ means that we sum over R-charge-squared of  the right-moving fermions and subtract the R-charge-squared of the left-moving fermions. Importantly, because this is the anomaly coefficient, it can be computed away from the critical point. 
This method has been used to compute the central charge in a number of examples, including ${\cal N}=(0,2)$ models \cite{evaed} and ${\cal N}=(2,2)$ Landau-Ginzburg theories \cite{lg}, Liouville theories \cite{hk} and non-Abelian gauge theories \cite{glop}\footnote{It is also simple to check that the this method can be used to reproduce the 
central charge of the Higgs branch CFT in $\ff$ theories. It is unable reproduce the Coulomb branch central charge because, as explained in \cite{wittenhiggs}, the appropriate right-moving R-symmetry is not visible in the UV gauge theory.}.

\para
As always, the subtlety in this calculation lies in the identification of the infra-red R-current. It is not necessarily the same as the UV R-symmetry because this may mix with any global Abelian flavour symmetries $F$ under the renormalisation group flow. The infra-red R-current is fixed simply by  requiring that it has no mixed anomalies with these flavour symmetries \cite{evaed},
\be {\rm Tr} \,FR=0\label{fr}\ee
This can be repackaged as a statement of c-extremisation \cite{us,bb}, in analogy to the story of a-maximisation in four dimensions \cite{a}

\para
How to apply the discussion above to our $\zf$ theory? Now the R-symmetries are no longer Abelian. Instead, they take the form $SU(2)_R^-\times SU(2)_R^+$ and their non-Abelian nature means that they cannot mix with 
any of the flavour symmetries. Instead, what is at stake is the identification of the infra-red $\zt$ sub-algebra. The R-current should be a linear combination of the two Cartan generators $R^\pm\subset su(2)_R^\pm$. 

\para
Viewing our gauge theory through $\zt$ eyes, we will see that both $R^+$ and $R^-$ are candidate R-symmetries, but only a specific linear combination obeys \eqn{fr}. This combination can then be used to calculate the central charge.

\subsection*{Global Symmetries}

We start with the global symmetries. It will be useful to gather together the various interactions of our ${\cal N}=(0,4)$ gauge theory. These are encoded in the $E$-terms and superpotentials. The former are
\be E_\Theta = [Y,\tilde{Y}]+\Phi'\tilde{\Phi}' \ \ ,\ \ 
 E_\Lambda = [Y,Z] \ \ ,\ \ 
 E_{\tilde{\Lambda}} = [Y,\tilde{Z}]\nn\ee
 \be 
E_\Gamma = Y\Phi\ \ ,\ \  E_{\tilde{\Gamma}} = -\tilde{\Phi}Y\ \ ,\ \
E_{\Gamma^\prime}=Z\Phi'\ \ ,\ \  E_{\tilde{\Gamma}{}^\prime}=-\tilde{\Phi}'Z\nn\ee
%
%
while the superpotential terms are
\be {\cal W}_\Theta = {\rm Tr}\ \Theta[Z,\tilde{Z}] + \tilde{\Phi}\Theta\Phi\ \ ,\ \ 
W_\Lambda = {\rm Tr}\ \Lambda[\tilde{Y},\tilde{Z}] \ \ ,\ \ {\cal W}_{\tilde\Lambda} = {\rm Tr}\ \tilde{\Lambda}[\tilde{Y},Z] \nn\ee
\be {\cal W}_\Gamma = \tilde{\Phi}\tilde{Y}\Gamma\ \ ,\ \ {\cal W}_{\tilde{\Gamma}}  =\tilde{\Gamma}\tilde{Y}\Phi\ \ ,\ \ {\cal W}_{\Gamma'} = \tilde{\Phi}'\tilde{Z}\Gamma'\ \ ,\ \ {\cal W}_{\tilde{\Gamma}{}^\prime} = \tilde{\Gamma}'\tilde{Z}\Phi'\nn\ee
Finally, we also have the couplings associated to the D5-D5\pprime\ strings
\be {\cal W}_\chi = \tilde{\Phi}\Phi'\ \ \ \ {\rm and}\ \ \ \ E_\chi = -\tilde{\Phi}'\Phi\nn\ee
%
%
%
%
%
%
We have already seen that the theory constructed in the previous section has an $SU(2)_L^-\times SU(2)_L^+$ global symmetry. We denote the Cartan subalgebra as $L^\pm \subset su(2)^\pm_L$. The vector multiplet fields have $L^\pm[\Upsilon] =L^\pm[\Theta]=0$. Similarly, the D5-D5\pprime\ strings are uncharged under these symmetries: $L^\pm[\chi]=0$. The charges of the remaining fields are
\begin{center}
\begin{tabular}{c|cccccccccccccc}
\ & $Y$ & $\tilde{Y}$ & $Z$ & $\tilde{Z}$ & $\Lambda$ & $\tilde{\Lambda}$ & $\Phi$ & $\tilde{\Phi}$ & $\Gamma$ & $\tilde{\Gamma}$ & $\Phi'$ & $\tilde{\Phi}'$ & $\Gamma'$ & $\tilde{\Gamma}{}^\prime$ \\ \hline
$L^-$ &  0 & 0 & +1 & $-1$ & +1 & $-1$ & 0 & 0 & 0 & 0 & 0 & 0 & +1 & +1  \\
$L^+$ & +1 & $-1$  & 0 & 0 & +1 & +1 & 0 & 0 & +1 & +1 & 0 & 0 & 0 & 0  \\
\end{tabular}
\end{center}
However, these are not the only global symmetries. There is a flavour symmetry which rotates the Fermi multiplet fields arising from the D5-D5\pprime \ strings. We call this $F_\chi$. All adjoint-valued multiplets have vanishing charge: $F_\chi[\Upsilon] = F_\chi[\Theta] = F_\chi[Y] = F_\chi[\tilde{Y}] = F_\chi[Z] = F_\chi[\tilde{Z}] = F_\chi[\Lambda] = F_\chi[\tilde{\Lambda}]=0$. The charges of the remaining fields are
\begin{center}
\begin{tabular}{c|ccccccccc}
\ &  $\Phi$ & $\tilde{\Phi}$ & $\Gamma$ & $\tilde{\Gamma}$ & $\Phi'$ & $\tilde{\Phi}'$ & $\Gamma'$ & $\tilde{\Gamma}{}^\prime$ & $\chi$ \\ \hline
$F_\chi$ &  +1 & $-1$ & +1 & $-1$ & $-1$ & +1 & $-1$ & +1 & $+2$   
\end{tabular}
\end{center}
Finally, there is one further $U(1)$ that is a flavour symmetry from the perspective of ${\cal N}=(0,2)$ superfields. (We shall see its origin from the ${\cal N}=(0,4)$ perspective shortly). We call this $U(1)$ action $V$; it  doesn't act on the D5-D5\pprime \ strings so $V[\chi] =0$. The charges of all the other fields are given by
\begin{center}
\begin{tabular}{c|ccccccccccccccc}
\ & $\Theta$ & $Y$ & $\tilde{Y}$ & $Z$ & $\tilde{Z}$ & $\Lambda$ & $\tilde{\Lambda}$ & $\Phi$ & $\tilde{\Phi}$ & $\Gamma$ & $\tilde{\Gamma}$ & $\Phi'$ & $\tilde{\Phi}'$ & $\Gamma'$ & $\tilde{\Gamma}{}^\prime$ \\ \hline
$V$ & 2 &  +1 & +1 & $-1$ & $-1$ & 0 & 0 & $-1$ & $-1$ & 0 & 0 & $+1$ & $+1$ & 0 & 0  \\
\end{tabular}
\end{center}
It is simple to check that all of these symmetries, $L^\pm$, $V$ and $F_\chi$, are compatible with the $E$-terms and superpotentials. 

\subsection*{R-Symmetry}

We now turn to the R-current. As we explained above, the $\zf$ theory enjoys an $SU(2)_R^-\times SU(2)_R^+$ R-symmetry. Because these are unable to mix with Abelian flavour symmetries, the appropriate $U(1)_R$ current $R$ must be a linear combination of the Cartan generators $R^\pm \subset su(2)_R^\pm$. 
As we described in Section \ref{dbranesec}, the matter Fermi multiplets $\Lambda$, $\tilde{\Lambda}$, $\Gamma$, $\tilde{\Gamma}$, $\Gamma'$ and $\tilde{\Gamma}'$ are all singlets under $SU(2)_R^+\times SU(2)_R^-$. The remaining superfields have R-charges given by
\begin{center}
\begin{tabular}{c|cccccccccc}
\ & $\Upsilon$ & $\Theta$ & $Y$ & $\tilde{Y}$ & $Z$ & $\tilde{Z}$ & $\Phi$ & $\tilde{\Phi}$ & $\Phi'$ & $\tilde{\Phi}'$ \\ \hline
$R^-$ &  1 & $-1$ & 0 & 0 & 1 & 1 & 1 & 1 & 0 & 0   \\
$R^+$ & 1 & 1 & 1 & 1 & 0 & 0 & 0 & 0 & 1 & 1  \\
\end{tabular}
\end{center}
The overall signs of these charges are fixed by requiring $R^\pm[\Upsilon]=+1$. It is simple to check that the charge assignments are consistent with the E-terms and superpotentials listed above. Each superpotential carries R-charge $R[{\cal W}]=+1$ while, for each Fermi multiplet  $\Psi$, the corresponding E-term has charge $R[E_\Psi] = R[\Psi] + 1$. 

\para
Both $R^-$ and $R^+$ are  candidates for the $\zt$ R-symmetry current. In fact, any linear combination of the form $R=a_-R^-+a_+R^+$ with $a_-+a_+=1$ is a candidate $R$-symmetry current. But which is the correct one? By construction, both $R^+$ and $R^-$ are orthogonal to $L^\pm$ and $F_\chi$. We are left only with the $U(1)$ symmetry $V$. However, while this is a flavour symmetry from the ${\cal N}=(0,2)$ perspective, it is not a flavour symmetry  of full ${\cal N}=(0,4)$ algebra. Indeed, it is simple to check that $V=R^+-R^-$. In particular, one can check that this combination acts equally on bosons and their right-handed fermionic partners in ${\cal N}=(0,2)$ chiral multiplets. Nonetheless, if we're looking for the relevant ${\cal N}=(0,2)$ R-symmetry current, we still require ${\rm Tr}\,VR=0$.
One finds that
\be {\rm Tr}\,VR^- &=&  -2Q_1 \qfm
\nn\\{\rm Tr}\, VR^+&=&  + 2 Q_1\qfp\nn\ee
(Note: in doing these calculations, it's important to remember that we're summing over the R-charges of the fermions. For left-moving fermions, this is the same as the R-charge of the Fermi multiplet. For right-moving fermions, which live in chiral multiplets, it is 
$R[\psi_+] = R[\Phi]-1$). We learn that the appropriate linear combination is\footnote{An Aside: One can also derive this result in a more pedestrian manner by forgetting that the $U(1)_R$ originated in a non-Abelian group and simply writing down the most general Abelian R-symmetry consistent with the constraints imposed by superpotentials and E-terms. One finds that there is a unique right-moving R-symmetry that is orthogonal to the flavour symmetries $L^\pm$, $V$ and $F_\chi$: it is given by \eqn{roar}.}
\be R = \frac{\qfp}{\qfp+\qfm}\,R^- + \frac{\qfm}{\qfp + \qfm}\,R^+\label{roar}\ee
where the normalisation is fixed by the requirement that $R[\Upsilon]=+1$ as befits the multiplet containing the field strength.

\subsection*{Computing the Central Charge}

With the R-charge assignments in hand, it is a simple matter to compute the central charge. The only non-vanishing contributions are

\be \hat{c}_R &=& Q_1^2\left(- R[\Upsilon]^2 - R[\Theta]^2 + (R[Y]-1)^2 + (R[\tilde{Y}]-1)^2 + (R[Z]-1)^2 + (R[\tilde{Z}]-1)^2\right) \nn\\ 
&& + Q_1\qfp\left((R[\Phi]-1)^2 + (R[\tilde{\Phi}]-1)^2 \right) + Q_1\qfm\left((R[\Phi']-1)^2 + (R[\tilde{\Phi}']-1)^2\right)\nn\ee
The contribution from the adjoint-valued fields cancel. We're left with the central charge
\be \hat{c}_R = 2Q_1\frac{\qfp\qfm}{\qfp+\qfm}\label{itworks}\ee
in agreement with the central charge of the supergravity solution \eqn{c}. 

\para
Given the D-brane origin of the $\zf$ gauge theory, together with the agreement of its central charge, it is natural to conjecture that the gauge theory flows in the infra-red to the large $\ff$ superconformal fixed point which is the holographic dual of $\adss$ supported by RR flux.
 However, the original gauge theory was chiral and we still have to account for the  $\qfp\qfm$ extra left-handed fermions $\chi_-$. It is plausible that these simply decouple, presumably after mixing with other left-moving fermions such as  $\psi_-$ and $\lambda_-$. A similar fate for these fermions in the absence of the D1-strings was suggested in \cite{ibrane}. If this is the case the low-energy physics is  described by the interacting large $\ff$ theory, together with a separate sector of left-moving fermions. 

\para
One may wonder if the conformal field theory with this central charge is associated to one of the branches of vacua that we saw in Section \ref{vacsec}. This cannot be. All branches of vacua have asymptotic regimes where $Y, \tilde{Y}\neq 0$ or $Z,\tilde{Z}\neq 0$. Yet, in such asymptotic regimes, where semi-classical analysis is trustworthy, a purely right-moving current like $R$ cannot act on scalar fields which have no decomposition into separate right-moving and left-moving pieces. This means that the central charge $\hat{c}_R$ is likely to be telling us about modes localised at the origin of the brane intersection. Needless to say, this is rather unusual behaviour; in two-dimensions,  wavefunctions typically spread over moduli  spaces unless branches are renormalised to infinite distance in field space. This issue seems likely to be at the heart of the  ${\bf S}^1$ vs ${\bf R}$ mystery  seen in the geometry and it would certainly be interesting to understand this better. Presumably, although localised, the CFT has singularities and the associated continuous spectrum above some threshold arising from the presence of the other branches \cite{sw}.

%
%

\section{Open Questions}\label{endsec}

We end with a discussion of some future directions. 

\para
First, and most obviously, we have only shown that the IR central charge of the ${\cal N}=(0,4)$ gauge theory coincides with that of string theory on $\adss$. Much more is known from the supergravity side, including the spectrum of BPS Kaluza-Klein states (to leading, linear order). If the proposal of this paper is correct, these should match the chiral primary operators that can be constructed from the  gauge theory degrees of freedom. Moving beyond, supergravity flows of the theory were studied in \cite{berg}; aspects of integrability of strings in $\adss$ has been explored, starting in \cite{bogdan}.

\para
The next set of questions are not directly related to the $\adss$ geometry, but instead concern results about related systems of D-branes.  The dynamics of the intersecting D5-branes in the absence of D1-strings was studied in detail in \cite{ibrane}  where a number of dramatic results were uncovered. Most notably, it was found that anomaly inflow \cite{michael} causes the chiral fermions to be pushed away from the $d=1+1$ dimensional intersection region which, to a low-energy observer, instead looks like it has $d=2+1$ dimensional Poincar\'e invariance and enhanced supersymmetry. 

\para
There should, presumably, be some hint of this behaviour in the $\zf$ gauge theory constructed in this paper. The enhanced supersymmetry and restoration of chiral symmetry seems to have a counterpart. But what of the enhanced Poincar\'e symmetry and the relocation of the chiral fermions?
Perhaps a lesson on how to proceed can be taken from studies of the D1-D5 system. There, many aspects of the Higgs branch can be understood by integrating out the hypermultiplets and focussing on the throat-like region that arises at the origin of the Coulomb branch \cite{wittenhiggs,ab}. This takes the form,
\be ds^2  = dr^2 + \frac{Q_5}{2}d\Omega_3^2 \nn\ee
with torsion $H=-Q_5d\Omega$. and a linear dilaton charge. For the $\zf$ gauge theory studied in this paper, a similar calculation gives rise to two such throat geometries, coupled through the left-moving chiral fermions $\chi_-$ and $\tilde{\chi}_-$ which mix with corresponding left-moving fermions associated to each throat. It would certainly be interesting to understand the interplay between this dynamics and the features of intersecting D5-branes discovered in  \cite{ibrane}.

\para
Finally, it was recently suggested that theories exhibiting large $\ff$ superconformal symmetry may be related to higher spin theories in AdS${}_3$ \cite{gg}. The gauge theoretic UV description of a class of these fixed points may help in  shedding light on this.

\section*{Acknowledgements}

My thanks to  Allan Adams, Ofer Aharony, Micha Berkooz, Chi-Ming Chang, Aristos Donos, Nick Dorey, Shu-Heng Shao, Kenny Wong and Xi Yin 
for many useful discussions and comments. I am supported by STFC and by the European Research Council under the European Union's Seventh Framework Programme (FP7/2007-2013), ERC Grant agreement STG 279943, Strongly Coupled Systems

\end{document}